\documentclass[conference]{IEEEtran}
\IEEEoverridecommandlockouts

\usepackage{cite}
\usepackage{amsmath,amssymb,amsfonts}
\usepackage{algorithmic}
\usepackage{graphicx}
\usepackage{textcomp}
\usepackage{xcolor}

\usepackage{circledsteps}
\usepackage{multirow}
\usepackage{booktabs} 

\def\BibTeX{{\rm B\kern-.05em{\sc i\kern-.025em b}\kern-.08em
    T\kern-.1667em\lower.7ex\hbox{E}\kern-.125emX}}

\newcommand{\circled}[1]{\Circled[fill color=black, inner color=white]{#1}}

\makeatletter 
\newcommand{\linebreakand}{%
  \end{@IEEEauthorhalign}
  \hfill\mbox{}\par
  \mbox{}\hfill\begin{@IEEEauthorhalign}
}
\makeatother 

\begin{document}

\title{ShardMeter: Sharded and Geo-Distributed Training Without the Guesswork}

\author{\IEEEauthorblockN{Tim Beringer$^{1}$, Patrick Diem$^{1}$, Felix Wolf$^{1}$, Arya Mazaheri$^{1,2}$}
\IEEEauthorblockA{$^{1}$Technical University of Darmstadt \\
$^{2}$PanocularAI \\
\{tim.beringer, patrick.diem, felix.wolf, arya.mazaheri\}@tu-darmstadt.de}
arya@panocular.ai

}

\maketitle
\begin{abstract}
Training large-scale AI models often outgrows a single data center, demanding sharded, multi-cluster, and decentralized training. However, the huge space of resource allocations makes exhaustive benchmarking and manual tuning impractical, while performance depends on tightly coupled factors like model size, GPU memory, batch size, bandwidth, and sharding strategy. We introduce ShardMeter, a lightweight analytical performance model that predicts the end‑to‑end runtime of transformer‑based workloads across arbitrary sharded, distributed, and even decentralized training. Given a model's characteristics and a target hardware topology, ShardMeter estimates per‑GPU and per‑island throughput, training cost, total wall‑clock time, and identifies performance bottlenecks. Our analysis reveals diminishing‑return regimes as island size increases, quantifies transitions between compute‑ and communication‑bound scaling, evaluates hyperparameter trade‑offs, and models cost–throughput for large‑scale decentralized training. ShardMeter exposes these insights to quickly explore the configuration space, choose near-optimal deployment plans, and avoid costly trial and error.
\end{abstract}


\section{Introduction}
\label{sec:introduction}

Recent advancements in large foundation models, which are typically based on transformer architectures~\cite{transformer}, have revolutionized numerous AI applications~\cite{zhao2023survey}, including natural language understanding and generation (e.g., machine translation and summarization)\cite{brown2020gpt3}, computer vision (e.g., image classification and zero-shot recognition)\cite{dosovitskiy2021vit,radford2021learning}, and multimodal reasoning (e.g., aligning text and images for retrieval and captioning)~\cite{radford2021learning,Li2024MultimodalAAA}. 
However, training such models demands substantial computational resources and energy. Traditionally, data parallelism has been one of the primary approaches in distributed training, but the sheer size of the model--sometimes exceeding hundreds of billions of parameters--necessitates other parallelization schemes, such as partitioning across multiple GPUs using techniques like fully sharded data parallelism (FSDP)~\cite{zhao2023pytorch}.
Consequently, FSDP has become the essential strategy among others for training medium‑ to large‑scale transformer models and a number of state‑of‑the‑art frameworks are now based on PyTorch's FSDP implementation~\cite{liang2025torchtitan,teknium2025hermes4technicalreport,jaghouar2024intellect1technicalreport,primeintellect2025prime-rl}.

Even within individual data centers, computational resources may be insufficient, requiring the coordination of multiple facilities through decentralized and federated learning (FL)~\cite{mcmahan2017communication}. 
Determining an efficient distributed training strategy is challenging because individual workloads scale differently with model architecture, dataset, and hardware, making it difficult to estimate the computational complexity of a training job~\cite{mccandlish2018}. 
Moreover, expanding a distributed-training setup to a sharded, geographically distributed setup adds further layers of complexity, introducing new potential bottlenecks related to network latency, data synchronization, and resource allocation. 
Therefore, determining an optimal training configuration remains a significant challenge, particularly for large language models (LLMs), which require substantial time and resources and are highly sensitive to suboptimal training parameters.

Within the context of decentralized training, critical choices include selecting appropriate data centers, the number of nodes, the number of GPUs per node, and an appropriate parallelization strategy, including a sharding level based on model size and available GPU memory. 
The performance of training workloads is shaped by a tightly coupled set of factors: model size, per-GPU memory capacity, batch size, GPU performance, network fabric, and the chosen parallelism scheme.
Together, they determine whether the workload is communication-bound or compute-bound, which significantly influences the scalability of individual training workloads.

As a result, developers often rely on ad hoc heuristics, empirical tuning, or costly trial‑and‑error experimentation to configure large‑scale training jobs. These approaches are time‑consuming, expensive, and brittle, as optimal configurations can vary substantially across models, datasets, and hardware environments~\cite{Qiao2020PolluxCCA}. In the absence of principled guidance, suboptimal choices may lead to severe underutilization of resources, excessive communication overheads, or even infeasible training runs due to memory constraints. This motivates the need for a predictive performance model that can reason about the interplay between computation, communication, and memory at scale, and that can estimate training efficiency and cost across a wide range of distributed and decentralized configurations before execution~\cite{strati2025sailor}.

However, existing performance models either target single‑GPU or conventional data‑parallel training, assume homogeneous hardware, and are limited to convolutional or regular distributed-data-parallelism workloads~\cite{trainmgpu2019,paleo,paradl,DNNPerf,beringer2024dissecting}.
Many approaches also require profiling a specific model before they can make any predictions, which prevents them from predicting unknown models~\cite{Habitat,strati2025sailor}.
Moreover, existing performance models cannot capture the complex, tightly-coupled inter-dependencies between communication and computation that arise in the widely-adopted, state-of-the-art FSDP and DiLoCo training for transformer models.

To address these challenges, we introduce ShardMeter, a lightweight analytical model that predicts end-to-end training time, pinpoints bottlenecks, and suggests near-cost-optimal deployment schedules. It quantifies the communication-compute trade-offs of a given configuration and predicts the scalability of training from a single-node, locally sharded setup to large, geographically distributed training workloads.

The main contributions of this paper are as follows:
\begin{itemize}
\itemsep0em 
    \item A lightweight analytical runtime-prediction model that captures the intertwined compute-communication dynamics of FSDP and decentralized training, with a median error of $<$15\%.
    \item A systematic bottleneck analysis that reveals diminishing return regimes, quantifies compute‑communication scaling transitions, and clarifies the impact of training hyperparameters.
    \item An integrated optimizer exhaustively explores heterogeneous GPU deployments, constructs a Pareto frontier of cost versus runtime, and can cut training expenses by up to 21\% while preserving near‑optimal throughput. 
\end{itemize}

\section{Background}
\label{sec:background}
In this section, we provide a brief overview of the principal distributed-training primitives that underpin modern LLM training, introduce the terminology used throughout the paper, and describe how sharding techniques and recent decentralized training are combined in practice.

\subsection{Deep-learning training fundamentals}

Training a deep neural network involves repeatedly feeding mini-batches of data through a forward pass, computing a loss, and applying a backward pass that yields gradients for the model parameters. 
To improve training throughput and mitigate hardware constraints imposed by large model sizes, a range of parallelization techniques has been introduced, including data, pipeline, context, tensor, and expert parallelism~\cite{ultrascaleplaybook}. This paper examines data parallelism in detail.

\paragraph{Distributed data parallelism (DDP)} The most popular parallelization strategy is DDP, where each GPU holds a replica of the full model and processes a distinct micro‑batch. After the backward pass, gradients are synchronized across all ranks using an all‑reduce collective. The latency of this operation is primarily determined by the network bandwidth of the intra-cluster fabric (e.g., NCCL + InfiniBand). Because DDP requires the full model parameter set to reside on each device, the approach is only applicable when the entire model fits into the memory of a single GPU, including space required for optimizer states, activation checkpoints, and micro-batches.

\paragraph{Fully-sharded data parallelism (FSDP)}
When model parameters no longer fit into a single GPU, FSDP exclusively partitions the model, the optimizer states, and the activation buffers across the participating ranks~\cite{zhao2023pytorch}. 

Before a forward pass, the required parameter shards are obtained with \texttt{all\_gather}, and after the backward pass, the gradients are combined with \texttt{reduce\_scatter}. Communication happens inside so-called islands (i.e., groups of nodes that communicate internally), typically within a node or over the fast local network. 
Thus, FSDP drastically reduces per‑GPU memory usage while adding extra communication within every training step.

\begin{figure}[t]
\centering
\includegraphics[width=\linewidth]{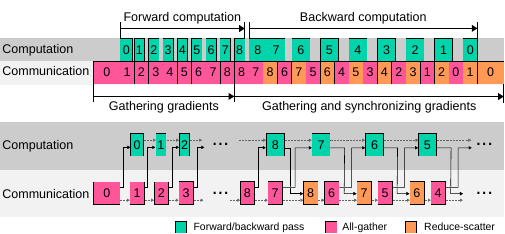}
\caption{(Top) Timeline of a single FSDP step showing overlapped communication and computation. (Bottom) Dependency graph of communication (all‑gather, reduce‑scatter) and compute (forward, backward) kernels. 
}
\label{fig:fsdp_dep}
\vskip -0.15in
\end{figure}

Figure~\ref{fig:fsdp_dep} shows the communication/computation pattern of a FSDP training step.
The top view shows the runtime of each operation that constitutes a full FSDP training step on a GPU, highlighting the overlap between communication and computation. Each compute operation usually consists of multiple GPU kernels, representing forward and backward pass operations.
In FSDP, the model is partitioned into several FSDP units, which are groups of layers used to lower memory usage. Rather than executing a single large \texttt{all\_gather} and \texttt{reduce\_scatter}, each unit performs its own forward and backward pass, with the corresponding communication. 
The bottom view depicts the explicit dependencies among the individual communication and computation primitives (i.e., the \texttt{all\_gather} operations before each corresponding forward and backward pass and the \texttt{reduce\_scatter} operation afterward).  
During gradient accumulation, \texttt{reduce\_scatter} is omitted, but \texttt{all\_gather} remains to assemble the sharded gradients.
Communication and computation can overlap, but at any instant only one of each may be active; therefore, whether a step is compute‑bound or communication‑bound depends on the relative speed of the hardware and the workload.

\subsection{Decentralized training}
Decentralized training extends the distributed-training strategy to multiple geographically dispersed training islands. 
Each island computes its gradients locally and only exchanges model updates after a predefined interval. This approach is crucial when the workload size or privacy requirements necessitate the use of distinct locations, but it also highlights communication as a significant bottleneck.

To address the communication bottleneck, the distributed low-communication training algorithm (DiLoCo)~\cite{douillard2023diloco,douillard2025streaming} organizes computing resources into isolated training islands, dividing the training process into two stages. The model is replicated within each island, and within these islands, the model is sharded across GPUs using FSDP. This ensures that collective operations take place over the fast intra-island network. After a configurable number of local steps, each island forms an outer model and computes a pseudo‑gradient, the gradient of the loss with respect to the local parameters prior to global reduction. 
These pseudo-gradients are then synchronized across islands with \texttt{all\_reduce} over the mainstream internet. In this way, DiLoCo leverages FSDP for efficient intra-island sharding, while dramatically reducing the volume of data that must traverse the slower inter-island network.

Recent extensions of the DiLoCo algorithm, such as 
Streaming DiLoCo~\cite{douillard2025streaming}, mitigate the cost of exchanging the full pseudo-gradient at a single synchronization point. 
Instead of communicating the entire pseudo-gradient at once, it partitions gradients into smaller units (fragments) that are sent incrementally.
Each fragment covers a configurable subset of layers-- users set granularity and the communication schedule. Interleaving fragment exchanges with computation spreads communication evenly across training iterations.
This design increases the total
number of communication events, but significantly reduces peak bandwidth requirements and alleviates synchronization-induced stalls.

\section{Approach}
\label{sec:approach}
We present ShardMeter, a performance model designed to analyze and predict the behavior of FSDP and decentralized training workloads.
ShardMeter estimates the per‑step training cost in both centralized and geographically distributed settings and automatically derives an efficient deployment plan for a given model architecture and target hardware platform. 
We employ a regression‑based model in ShardMeter to capture the nuanced, nonlinear interplay between communication and computation in FSDP‑ and DiLoCo‑based training. Starting from regression models that predict the cost of each communication and compute operation, ShardMeter builds execution graphs that encode their dependencies and possible overlaps, enabling critical‑path analysis of the training pipeline. 
While ShardMeter's simulation of FSDP and Streaming DiLoCo is inspired by the implementation in TorchTitan~\cite{liang2025torchtitan}, it targets a generic FSDP and Streaming DiLoCo protocol and does not depend on any particular PyTorch framework.

\begin{figure}[t]
  \centering
  \includegraphics[width=\linewidth]{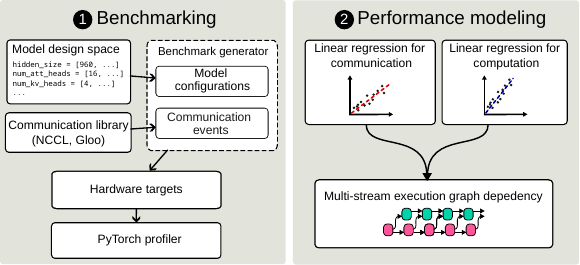}
  \caption{ShardMeter's workflow: \circled{1} benchmark to collect regression data; \circled{2} build a performance model that respects execution‑graph dependencies. }
  \label{fig:shardmeter_overview}
  \vskip -0.15in
\end{figure}

As depicted in Figure~\ref{fig:shardmeter_overview}, the modeling process proceeds in multiple stages. First, we collect enough benchmarking data of various model configurations. Then, ShardMeter models individual logical compute and communication operations, fitting separate regression models for each operation type. Finally, it composes these operations into execution graphs that capture the higher‑level structure of FSDP and decentralized training schemes (e.g., DiLoCo), enabling end‑to‑end performance prediction and configuration optimization. The following sections describe each stage and its components in detail. %

\subsection{Model and hardware benchmarking}
To train a regression model that can reliably predict the training‑workload characteristics of any AI model, we must collect benchmark measurements from a large, representative sample of the full design space, spanning the entire range of possible model‑architecture parameters.

These data points must capture diverse computational and execution characteristics through detailed profiling. However, generating such a dataset is challenging due to the breadth of the design space, which includes numerous AI models with varying hyperparameters, as well as significant structural overlap among models. Moreover, each data point is hardware‑specific, requiring remeasurement when the model is executed on new hardware.
To address this challenge, we systematically generate representative workloads using a benchmark generator and collect fine‑grained execution traces for performance modeling. These traces are generated by exploring a predefined model‑design space. 

\subsubsection*{Benchmark generation}
To collect the required data points for populating the performance model that predicts the latency of individual communication and computation primitives, we conduct a series of profiling experiments on the target hardware platform. 
Instead of running different transformer models, we construct a parameterized catalog of transformer configurations that captures the diversity of modern language-model architectures. Table~\ref{tab:benchmark} lists the parameters we considered throughout our experiments along with their corresponding values. We select the boundaries based on values observed across different LLMs and heuristically select up to 200 configurations for our benchmark. Increasing the number of datapoints can modestly improve accuracy, though it does lengthen the benchmarking process.

\begin{table}[t]
\setlength{\abovecaptionskip}{-0.1in}
\caption{Static model parameters and typical hyper‑parameter values for benchmarking; batch size is increased until an out-of-memory error occurs. }
\label{tab:benchmark} 
\begin{center}
\begin{small}
\begin{tabular}{ll}
\toprule
Prarameter & Range \\
\midrule
Number of attention heads & $[16,32]$ \\
Hidden size & $[960, 4096]$ \\
Number of hidden layers & $[16,32]$ \\
Intermediate size & $[2560, 24576]$ \\
Number of key value heads & $[4, 8]$ \\
Batch size per GPU & $[1, \infty)$ \\
Number of GPUs & $[1, 16]$ \\
\bottomrule
\end{tabular}
\end{small}
\end{center}
\vskip -0.2in
\end{table}

We also measure the runtime of all collective operations, namely \texttt{all\_gather}, \texttt{reduce\_scatter} used within the training, and \texttt{all\_reduce} in decentralized scenarios. While the communication operations appear as a single GPU kernel in the profiling traces, a forward and backward pass typically consists of dozens of smaller kernels. To determine the total duration of the logical forward and backward phases, we group all kernels that belong to the same phase --- identified by their naming patterns and execution order --- and sum their individual runtimes.

\subsubsection*{Runtime profiling}
To collect raw runtime data for our performance model, we execute a short training loop on the target hardware with the PyTorch profiler enabled for each model in the predefined DNN design space -- that is, for each possible combination of the selected parameters as defined in the benchmark.  The profiler writes an execution trace file that records every GPU kernel launched during each training step, together with its start and end timestamps, device/stream identifiers, and the Python call stack that triggered it. 
By parsing the call stack, we can associate each kernel with the high‑level forward or backward operation that emitted it; the runtimes of all kernels belonging to the same operation are then summed to obtain the total execution time of that logical operation. The final dataset, forward and backward runtimes paired with the static parameters that produced them, serves as the input for the regression-based performance model.

\subsection{Communication and computation modeling}

A training workload can be described in terms of its constituent communication and computation components. 
ShardMeter uses this concept to represent and analyze the actual workload running on the hardware. In our model, we abstract GPU execution at a coarser granularity, focusing on the aggregate computational workload rather than individual GPU kernels, where a single forward or backward pass---either for an entire model or for a transformer block under FSDP---can invoke dozens of GPU kernels (e.g., SGEMMs and data movement operations). For the purpose of end‑to‑end runtime prediction, we therefore model each forward or backward pass as a single computational unit. 

In this section, we describe how we model the cost of both communication and computation primitives. We adopt a linear regression approach to fit the coefficients of a simple base function that predicts runtime from a set of static parameters.
The model is trained on data points obtained from benchmarking. By fitting a linear regression model to these points, we derive a compact, analytically tractable cost model that can be quickly evaluated during scheduling. We chose linear regression because it is computationally inexpensive and enables the exploration and evaluation of a large design space while still representing the complex, nonlinear interactions among variables. 
In the following, we will present the base functions that we use for communication and computation.

\subsubsection*{Communication modeling}
We model the impact of latency and bandwidth as sets of regression coefficients rather than single scalars to capture the two fundamental limitations of communication. 
Each element of the latency set $L$ corresponds to the one-way delay along a particular GPU-to-GPU communication path (e.g., PCIe, SXM, InfiniBand, or Ethernet). Analogously, each element of the bandwidth set $B$ represents the data‑transfer rate of that same path.
When we compute the communication time, we use the largest latency among the selected paths, since in ring-based collectives the overall communication time is limited by the slowest hop and the narrowest bandwidth; the overall throughput cannot exceed the capacity of the narrowest link. 
Altogether, we use the following model to predict the communication time:
\begin{equation*}
 T_{sync} = (P-1) * (\max (L) + \min(B) * D/P),
\end{equation*}
\noindent where $D$ is the size of the payload, and $P$ is the number of ranks involved in communication (across all participating nodes).
We tune the coefficients $L$ and $B$ for every interconnect and use the slowest coefficients from interconnects used in a communication. 

\subsubsection*{Computation modeling}

Similar to the base function for communication, we leverage the inherent architectural features of the DNN to estimate the computational cost of the forward and backward passes.
The primary motivation is to develop a generalizable model that predicts the computational time of previously unseen transformer models using only their static, extractable parameters. This approach eliminates the need to execute or profile new models on specific hardware, enabling runtime prediction without requiring further model finetuning. 

To develop the regression-based model, we use the data points obtained from the benchmarks and the parameters specified in the model design space. ShardMeter automatically tests different parameter combinations and selects those that contribute most to predicting computational time.
These parameters are static (i.e., they can be extracted from the model architecture) and independent of runtime conditions (e.g., hardware or network).

Using the selected features, we use linear regression to construct a model that predicts computation time $ T_{fwd/bwd}$. 
\begin{equation*}
T_{fwd/bwd} = \sum_{f\in F} c_f*f ,
\end{equation*}
\noindent where $F$ is a set of features and $c_f$ is the tunable coefficient of a feature $f$. A higher coefficient for a specific parameter suggests that increasing that parameter in the architecture disproportionately increases computational cost, helping identify bottlenecks in model design (e.g., optimizing for memory efficiency vs. speed) and whether they are specific to a given hardware. Because each parameter independently scales the dimensions of the underlying GPU kernels, the relationship between a parameter and the total cost is approximately linear; therefore, we model the cost as an additive sum of coefficient‑feature products and omit interaction terms.

\subsection{Multi-stream execution graph construction}
\label{sec:task_graphs}
Similar to trace outputs that profiling tools (e.g., NVIDIA Nsight Systems) generate for visualizing the timeline of CPU/GPU operations and bottleneck analysis, we aim to leverage performance models from computational and communication analysis to quantify runtime and dependencies. 
This approach enables us to model inter- and intra-stream dependencies, identify bottlenecks, and analyze how compute-communication overlap affects overall training efficiency. 
To capture the complex interactions between computation and communication in distributed training, we construct task graphs that represent the sequence of compute and communication operations across multiple concurrent execution streams (Figure~\ref{fig:fsdp_dep}).

Each task graph is divided into multiple streams, where tasks within a single stream execute sequentially. 
However, tasks across streams can run concurrently, provided that their inter-stream dependencies are satisfied. 
This structure reflects the parallelism observed in execution trace files, where compute and communication operations are often interleaved. 
Within a single stream, tasks are arranged in a linear sequence, with each task depending on the completion of the previous one.
Tasks across streams may require synchronization. For instance, within FSDP training, 
an \texttt{all\_gather} operation (i.e., collecting the sharded gradients from all ranks) must be completed before the corresponding 
forward/backward pass can begin in the compute stream. Similarly, a reduce-scatter (aggregating gradients) must wait for the compute block to finish before initiating the next communication step. These dependencies are modeled as synchronization edges between tasks in different streams.

This way, the task graph explicitly captures how compute and communication operations overlap in time, depending on the training strategy. 
After building a task graph that represents a training step, we assign a runtime function derived from the computational and communication performance models developed earlier in this section.
By inserting these runtime functions into the task graph, we can simulate the end-to-end training step and analyze how dependencies and resource contention affect performance.
For example, a compute-bound task (e.g., a large forward pass) may dominate the runtime, 
while a communication-bound task (e.g., a slow \texttt{all\_gather}) may create bottlenecks. 
The task graph allows us to identify critical paths (longest dependency chains) and optimize for 
compute-communication overlap by adjusting task scheduling or resource allocation.

\subsubsection{Execution graph for decentralized training}

For synchronous decentralized training methods such as DiLoCo, which includes a blocking, synchronous gradient exchange between all participating training islands, we add a second level on top of the FSDP-step execution graph. This second level extends the model to a global, decentralized training task graph, where each task corresponds to a nested FSDP task graph (i.e., a training island). This global task graph models the inter-island communication required for decentralized training, including aggregating model parameter updates across islands and synchronizing training steps among them.
In this higher-level graph, each stream corresponds to a concurrent execution of a training island (i.e., a local FSDP task graph).
Tasks at this level are nested within their respective local execution graphs, capturing the runtime of the entire training step, including local computation and communication.  
Inter-island communication tasks (e.g., outer step) are added as new tasks, with dependencies on the completion of nested local task graphs.

For asynchronous methods, such as Streaming DiLoCo, where outer-step synchronization is performed asynchronously, we extend the FSDP-step execution graph for another stream representing the outer-step communication. For Streaming DiLoCo, fragments are sent asynchronously during training, so we initiate a communication task once the fragment's corresponding gradients are computed. Additionally, we add a dependency that requires synchronizing the fragment before it is used in an upcoming iteration, as defined by the user's delay. This way, if the outer step takes too long, the training will pause. We defer a detailed analysis of this behavior to Section~\ref{sec:results}.

\subsection{Workload scheduling and deployment optimization}
ShardMeter can both predict the per‑step cost of FSDP training across centralized and decentralized scenarios and generate an optimal deployment plan for a given training job and hardware platform. The minimum input consists of the target model and a global batch size. Additionally, more constraints can be set, such as the maximum number of islands and a hard limit on the total number of GPUs. Given a fixed underlying hardware setup, ShardMeter explores the space of feasible configurations through an incremental search. First, for each node type, it enumerates all admissible island configurations $(N, G)$, where $N$ is the number of nodes in the island and $G$ is the number of GPUs per node. For each $(N, G)$, it estimates the largest micro‑batch size that can be accommodated while fully utilizing the GPUs, thereby minimizing the total number of training steps required. Using the performance model, it then simulates the island's throughput and monetary cost when run with this maximal micro‑batch size; these estimates provide a coarse overview of its throughput.

Next, ShardMeter evaluates all admissible island combinations that satisfy the specified constraints. The global batch size is initially apportioned among the islands in proportion to the relative throughputs obtained in the first stage. An iterative refinement then adjusts the per‑island batch sizes: for each proposed allocation, ShardMeter re‑simulates the runtime (again via the performance model) and enforces the global batch‑size target while guaranteeing that each island respects its minimum step size. This minimum batch step size is a function of the island’s GPU count and its micro‑batch size. The search terminates when a configuration that either (1) minimizes runtime, (2) minimizes cost, or (3) achieves a user‑specified trade‑off between the two is found. In this way, ShardMeter delivers a near-optimal deployment plan with respect to the chosen objective while honoring all user‑imposed constraints.

\section{Experimental results}
\label{sec:results}
In this section, we assess the prediction accuracy and effectiveness of ShardMeter. We start by validating its accuracy through benchmarking of the underlying compute and communication components. 
Following this, we evaluate ShardMeter's predictive capabilities across various levels, from individual forward/backward kernels to single-island and multiple-island training strategies. Additionally, we examine the performance model in the context of decentralized training, using variants of the DiLoCo algorithm.

\subsection{Experimental setup}

\subsubsection*{Hardware setup and benchmarks}

All single-node experiments were performed on a workstation comprising two Intel Xeon Gold 5318Y CPUs (24 cores in total) and four NVIDIA A100 GPUs (each equipped with 80 GB HBM2). 
Distributed trainings were conducted on an HPC cluster, where each node is powered by a pair of AMD EPYC 7402 CPUs and equipped with four NVIDIA A100 GPUs. Every compute node also incorporates four HDR‑200 InfiniBand network cards for inter‑node communication.

\subsubsection*{Dataset} 
The computation dataset was generated by repeatedly instantiating a transformer model while systematically varying a set of key hyperparameters (e.g., hidden size, number of attention heads, and key/value head count). 
For each configuration, we ran ten training iterations and extracted the execution times from the trace logs generated by the PyTorch Profiler~\cite{pytorch_profiler}. The extracted information, along with the training parameters, was then used to train the regression model. By sampling a broader region of the design space, the resulting performance model, as expected, attains greater statistical robustness.
To collect the communication dataset, we benchmark the communication time of collective operations in the communication library (e.g., NCCL) between various components (within a node and between nodes) for payload sizes ranging from 1 to 150 MiB.

\subsection{Validation of the performance model}

We evaluate ShardMeter's accuracy using the mean absolute percentage error (MAPE) and, since no other performance model exists for fully‑sharded data parallelism, directly compare its predictions with the measured runtime on the target hardware.

\subsubsection*{Communication modeling accuracy}

To analyze the accuracy of communication performance modeling with ShardMeter, we validate the ring-based collective operations that underpin distributed training, namely \texttt{all\_gather}, \texttt{reduce\_scatter}, and \texttt{all\_reduce}.  
Within \texttt{all\_gather} operations, we achieved a MAPE of 14.99\%, indicating reasonable accuracy for this operation. Within \texttt{reduce\_scatter} operation, we achieved a MAPE of 13.1\%. 
\texttt{all\_reduce} demonstrated the highest accuracy, with a MAPE of 2.14\%, confirming the model's precision in predicting the most complex collective operation used to synchronize the gradients between training islands.
These results validate the reliability of our communication performance model. The low MAPE values for \texttt{reduce\_scatter} and \texttt{all\_reduce} operations suggest that the model effectively captures the interplay of network topology, data size, and synchronization overhead.

\subsubsection*{Computation modeling accuracy}
We evaluated the forward and backward pass prediction time and accuracy across a range of transformer model architectures and size regimes: SmoLM 360M~\cite{allal2024SmolLM}, Llama 3.2 1B~\cite{grattafiori2024llama3herdmodels}, Phi-1.5 1.3B~\cite{textbooks2}, Gemma 2B~\cite{gemmateam2024gemmaopenmodelsbased}, and Yi 6B~\cite{ai2025yiopenfoundationmodels}. 
The resulting MAPE error rates ranged from 5\% to 29\%, with the median error for every operation and target device staying below 13\%. Table~\ref{sample-table} provides a detailed comparison of prediction results for the forward and backward pass across different batch sizes.

\subsubsection*{End-to-end training step accuracy}
The communication and computation primitives form the basis of the execution graphs we use to simulate fully‑sharded and decentralized training (Section~\ref{sec:task_graphs}). An FSDP step can be compute‑bound, communication‑bound, or a mix of both, depending on the factors discussed in Section (Section~\ref{sec:bottleneck_discovery}). Accordingly, its end‑to‑end prediction accuracy aligns more closely with either the compute‑error rate or the communication‑error rate, which is why we reported the error rates for both. In Table~\ref{sample-table}, forward and backward error rates are aggregated over all batch sizes tested. For the end‑to‑end runtime numbers, we use the largest batch that fits in memory on both platforms. All models exhibit error rates below 13\%, demonstrating the robustness of our performance model.

\begin{table}[t]
\setlength{\abovecaptionskip}{-0.1in}
\caption{Mean absolute percentage error (MAPE) – Cross-validated prediction accuracy across varying transformer models.\\Fwd: Forward pass, Bwd: Backward pass, Step: One training step}
\label{sample-table}
\vskip -0.15in
\begin{center}
\begin{small}
\resizebox{\columnwidth}{!}{
\begin{tabular}{llrrrrrr}
\toprule
\multicolumn{2}{c}{} & \multicolumn{3}{c}{A100} & \multicolumn{3}{c}{H200}\\ \cmidrule(rl){3-5}\cmidrule(rl){6-8}
Model & Size & Fwd  & Bwd & Step &  Fwd  & Bwd & Step \\
\midrule
SmolLM & 360M & 5.29 & 10.44 & 4.03 & 15.22 & 28.74 & 4.83 \\
Llama 3.2 & 1B & 5.09 & 10.13 & 8.68 & 8.44 & 6.97 & 6.88 \\
Phi-1.5 & 1.3B & 5.95 & 8.12 & 4.53 & 8.1 & 8.15 & 0.62 \\
Gemma & 2B & 5.45 & 4.15 & 12.34 & 12.64 & 6.56 & 8.98 \\
Yi-1.5 & 6B & 22.3 & 7.09 & 2.51 & 16.56 & 12.36 & 7.23 \\

\bottomrule
\end{tabular}
}
\end{small}
\end{center}
\vskip -0.2in
\end{table}

\subsection{Performance bottleneck discovery}
\label{sec:bottleneck_discovery}
FSDP training method reduces memory consumption by distributing model parameters across ranks. However, this memory-saving strategy introduces additional communication overhead. This increased communication alters the scaling behavior of training with respect to hyperparameter changes. Consequently, an FSDP training can be either compute-bound, communication-bound, or a mix of both, where different parts of each training step exhibit varying dependencies on computation and communication.
Understanding this transition is crucial because it identifies the training's specific bottlenecks and models its non-linear scalability. The exact point at which this shift occurs is influenced by a combination of factors, including hardware capabilities, model architecture, hyperparameter settings, and system configuration. ShardMeter can detect and report such intricate interdependencies to the user.

\subsubsection*{Batch size impact}
At small batch sizes, communication overhead dominates the system's performance. Frequent synchronization of gradients across devices, such as through \texttt{all\_gather} and \texttt{reduce\_scatter} operations, becomes the primary bottleneck. As the batch size increases, the system gradually transitions to a compute-bound state, where the time spent on compute-intensive operations, such as matrix multiplications and activation computations, becomes the dominant factor.

\begin{figure*}[t]
\centering
\includegraphics[width=\linewidth]{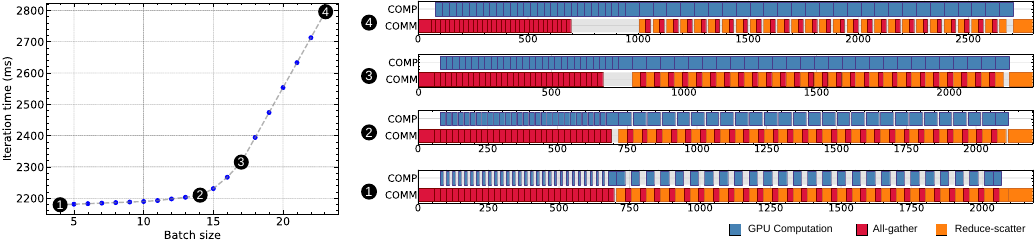}
\caption{Impact of batch size on compute idle time and overall throughput as simulated by ShardMeter.}\label{fig:graph_idle}
\vskip -0.2in
\end{figure*}

Figure~\ref{fig:graph_idle} illustrates the training step runtime of Gemma, demonstrating how the scaling behavior shifts between compute-bound and communication-bound regimes as the batch size varies. The accompanying dependence graph visualization on the right, generated by ShardMeter, highlights this behavior by showing gaps in the compute stream at low batch sizes. These gaps indicate periods where the compute pipeline is idle, waiting for communication operations to complete. As the batch size grows, compute-related gaps decrease, while communication-related gaps become more prominent.

\subsubsection*{Model size impact}
Large transformer models demand significant memory to store parameters, gradients, and activations. Due to their extensive use of dense matrix multiplications (SGEMMs) and other computationally intensive operations, these models often become compute-bound even at relatively small batch sizes. Hence, they benefit greatly from increased parallelism and optimized memory access patterns. In contrast, smaller models perform fewer computations per step and may remain communication-bound across a broader range of batch sizes. Moreover, the larger memory footprint of large models limits the maximum feasible batch size, potentially reducing scaling efficiency, especially in data-parallel training scenarios.

\begin{figure}[t]
\centering
\includegraphics[width=.9\linewidth]{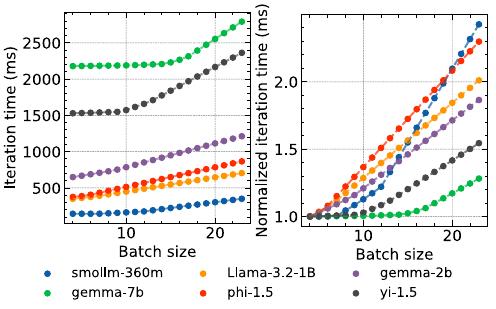}
\caption{Iteration time of individual models as a function of the device micro-batch size-- actual iteration time (left) and the relative increase  (right).}\label{fig:scalability_merged}
\end{figure}

Figure~\ref{fig:scalability_merged} (left) depicts how the iteration time (training step duration) varies with the device batch size for different models. The plots reveal a non-linear relationship between batch size and runtime as the system transitions from being communication-bound to compute-bound at a critical batch size. Understanding this transition is essential to understanding the performance characteristics of distributed training and to optimizing resource allocation.
Figure~\ref{fig:scalability_merged} (right) shows that the rate at which iteration time increases with batch size is not proportional to the model size. Larger models may exhibit shallower scaling in the communication-bound regime, whereas smaller models may show steeper increases. For instance, the SmolLM-360M model experiences a faster increase in runtime than larger models once it enters the compute-bound regime.

\subsubsection*{Hardware impact}
Apart from model size, GPU memory limitations (i.e., VRAM) can significantly restrict the maximum achievable batch size, even when the GPU offers high computational throughput. As available device memory decreases, the per-device batch size must be reduced accordingly. To maintain a fixed global batch size, this necessitates an increase in the number of gradient accumulation steps. In distributed data parallel (DDP) training, gradient accumulation typically incurs minimal overhead since gradients are communicated only during the final synchronization step.
In contrast, FSDP training incurs considerable overhead during gradient accumulation due to the need to reassemble parameter shards via \texttt{all\_gather} operations at every forward and backward pass. Therefore, for a fixed global batch size, increasing the per-GPU batch size directly reduces communication overhead in FSDP by decreasing the number of training steps required to achieve that global batch size. 
This effect is particularly significant for large models, where activations consume a substantial portion of GPU memory, leaving limited capacity for batch-dependent data and, thus, enforcing smaller per-device batch sizes.

Additionally, the performance of interconnects (e.g., NVLink vs. PCIe) is crucial. Faster interconnects reduce synchronization latency, enabling the system to shift towards compute-bound behavior at smaller batch sizes. The training island configuration also influences performance; if inter-node communication occurs in every training iteration (as in FSDP), communication overhead dominates even at moderate batch sizes. Conversely, if communication is limited to outer optimization steps in DiLoCo-style training, the system may remain compute-bound.

\begin{figure}[t]

\centering
\vskip -0.1in
\includegraphics[width=\linewidth]{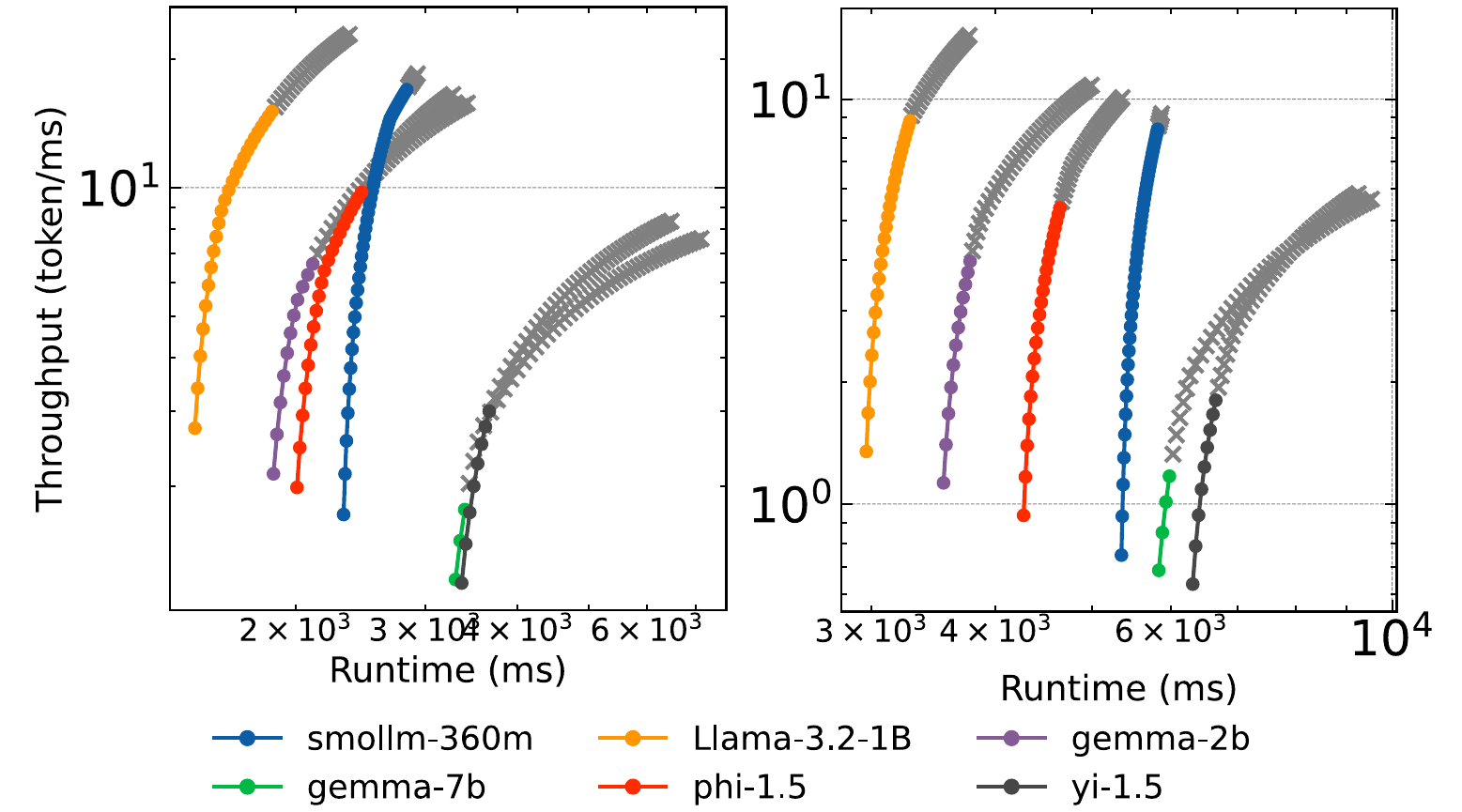}
\caption{Throughput and runtime trade-off estimation for different-sized models and different batch sizes on 8 GPUs (left) and 16 GPUs (right). Grey marker indicates batch sizes that are not supported on the hardware used.}\label{fig:throughput-tradeoff}
\vskip -0.1in
\end{figure}

\begin{figure*}[t]

\centering
\includegraphics[width=\linewidth]{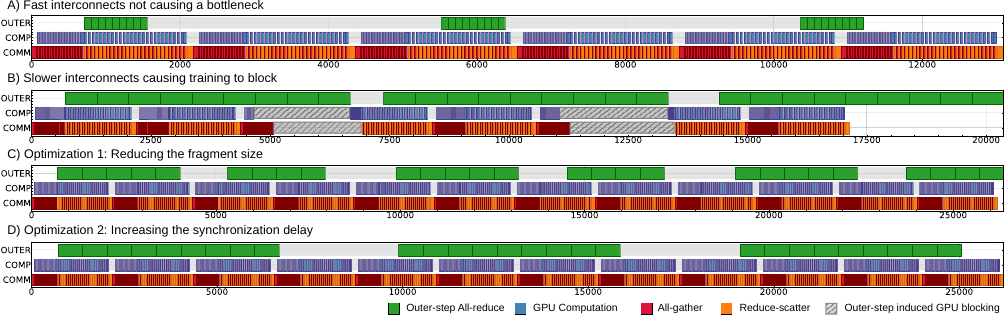}
  \caption{Traces of Streaming DiLoCo simulated using ShardMeter. Different rows within each configuration represent a stream. Red is all-gather, orange reduce-scatter, blue computation of forward and backward pass, and green all-reduce of the outer step.} 
  \label{fig:streaming_diloco_exp}
  \vskip -0.15in
\end{figure*}

Figure~\ref{fig:throughput-tradeoff} illustrates the trade-off between throughput and runtime for several transformer models across a range of batch sizes and two GPU configurations (i.e., 8 and 16 GPUs). For some models, throughput increases sharply at small batch sizes with only a marginal increase in runtime. This behavior occurs in batch-size regimes where training is communication-bound, meaning that increasing the amount of computation per step improves hardware utilization without substantially increasing end-to-end runtime.

Conversely, smaller models tend to become compute-bound at much lower batch sizes and therefore do not exhibit this steep initial throughput gain. Once computation dominates, further increases in batch size lead to proportional increases in runtime, resulting in diminishing returns on throughput. The batch size at which a model transitions from compute-bound to communication-bound also depends on the number of GPUs. 
Increasing the number of GPUs shifts this transition point due to the higher frequency and volume of communication introduced by ring-based parameter and gradient synchronization.

A key advantage of these estimates is their ability to explore batch sizes that exceed the VRAM capacity of the evaluated hardware. This enables us to infer how performance would scale on the same GPUs with larger memory budgets, providing actionable insights for future hardware procurements. Due to the communication overhead associated with gradient accumulation in FSDP, simply increasing the number of gradient accumulation steps would alter the scaling behavior. More importantly, the results highlight how VRAM constraints can severely limit achievable throughput, even when additional computational and communication capacity is available. 

\subsection{Simulating communication bottlenecks in asynchronous decentralized training}
Within the context of decentralized training (see Section~\ref{sec:background} for details), various synchronization strategies exist for managing gradient updates between training islands in the outer step. Blocking strategies, such as the original DiLoCo~\cite{douillard2023diloco}, where all participating islands pause at the outer step, simplify performance modeling. They only require calculating individual island runtimes to identify potential differences in completion time and idle periods. 
In the more advanced version of DiLoCo, namely Streaming DiLoCo~\cite{douillard2025streaming}, users specify a training step delay before gradients are required, and gradients are transmitted in fragments of a defined size, with one fragment sent every $i$ delay steps. This design introduces further potential bottlenecks, as the asynchronous outer-step communication may progress more slowly than the concurrent training. 
ShardMeter can simulate training execution under different hardware and hyperparameter settings to detect bottlenecks and propose solutions. 
Specifically, ShardMeter can determine the minimum number of inner steps required to complete a Streaming DiLoCo outer step within a given wall-time budget. This ensures the outer step remains synchronized with training progress, minimizing adverse effects on convergence and preventing the asynchronous outer step from becoming a performance bottleneck.

Figure~\ref{fig:streaming_diloco_exp} illustrates the simulated trace graphs representing different configurations, generated by ShardMeter. Configuration \circled{A} visualizes the compute and communication patterns of a typical Streaming DiLoCo setup, showing six training steps, including forward and backward passes, with an outer step occurring every two iterations. Crucially, these outer steps do not introduce any noticeable slowdown in the training process.
To explore potential issues, configuration \circled{B} artificially increases communication time to simulate a slower network. This causes the outer step to take longer than the training requires for the next blocking-receive operation, forcing the training to halt and wait, significantly increasing overall training time. There exist two potential solutions to mitigate this issue:
Increasing the number of fragments, as shown in \circled{C}, reduces the communication burden per outer step, eliminating the communication bottleneck. However, this increases the number of iterations required to fully exchange the model among all ranks. In this example, doubling the number of fragments increased the number of iterations from 6 to 12, potentially impacting the loss.
Alternatively, increasing the step size, for example, from 2 to 3, as illustrated in \circled{D}, extends the time interval until the results are needed, leaving more buffer for the outer step. This potential solution also increases the number of iterations required to fully exchange gradient updates. 

If synchronization is too infrequent, the replicas can drift apart, so a balanced inner‑step delay must be chosen~\cite{douillard2025streaming}. ShardMeter estimates the minimum viable inner‑step delay and the appropriate number of fragments by enforcing a lower bound on the time between asynchronous outer‑step communication and the blocking receive of the inner step. This ensures that training is not delayed while waiting for the outer step. 

ShardMeter selects the fewest fragments and the shortest training step delay to minimize the iteration delay until the pseudo-gradient is fully exchanged. As explained in detail in Section~\ref{sec:bottleneck_discovery}, ShardMeter accounts for all relevant factors, including model size, training configuration, and hardware characteristics, to compute a near‑optimal selection. Moreover, if any of these parameters change during training, ShardMeter can re‑evaluate the configuration and produce a new setting that avoids blocking the training.

\section{Case study: Modeling decentralized training}
\label{sec:casestudy}
Modeling the performance of training workloads is essential in a wide range of contexts, including the efficient scheduling of large-scale AI jobs. With the rapid growth in model sizes and the increasing adoption of decentralized training, having prior estimates of training time and cost has become more critical than ever. Consequently, identifying an efficient training plan requires a systematic and well-informed optimization process.

In this section, we demonstrate how ShardMeter evaluates and selects near-optimal deployment configurations for a specific transformer model training across heterogeneous GPU clusters under realistic hardware and cost constraints.
We selected the Llama-8B model with a fixed global batch size of 2 million tokens and defined a total of ten training islands. We restricted the number of training islands to a maximum of ten, following prior findings that larger island counts can negatively affect training stability~\cite{jaghouar2024intellect}. Decentralized training was simulated in a DiLoCo‑style setting on two heterogeneous clusters: Cluster A (10 nodes $\times$ 8 H200 GPUs, \$3.59 per GPU-hour) and Cluster B (10 nodes $\times$ 8 A100 GPUs, \$1.19 per GPU-hour). For the cost model, we used market prices sourced online~\cite{gpu_prices}.

\begin{figure}[t]
\centering
\includegraphics[width=\linewidth]{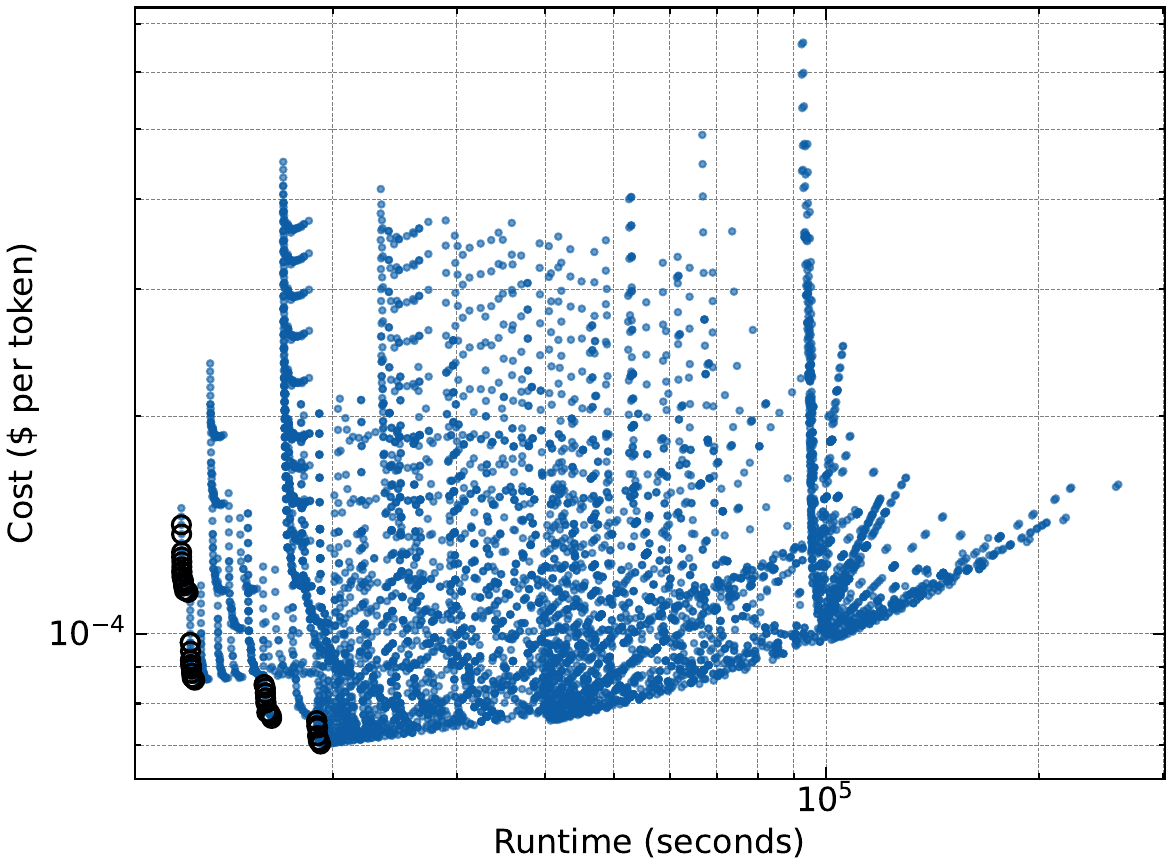}
\caption{All feasible deployment configurations, showing the trade-off between monetary cost and wall-clock runtime for a Llama-8B training job.}\label{fig:deployment}
\vskip -0.15in
\end{figure}

\begin{figure}[t]
\centering
\includegraphics[width=\linewidth]{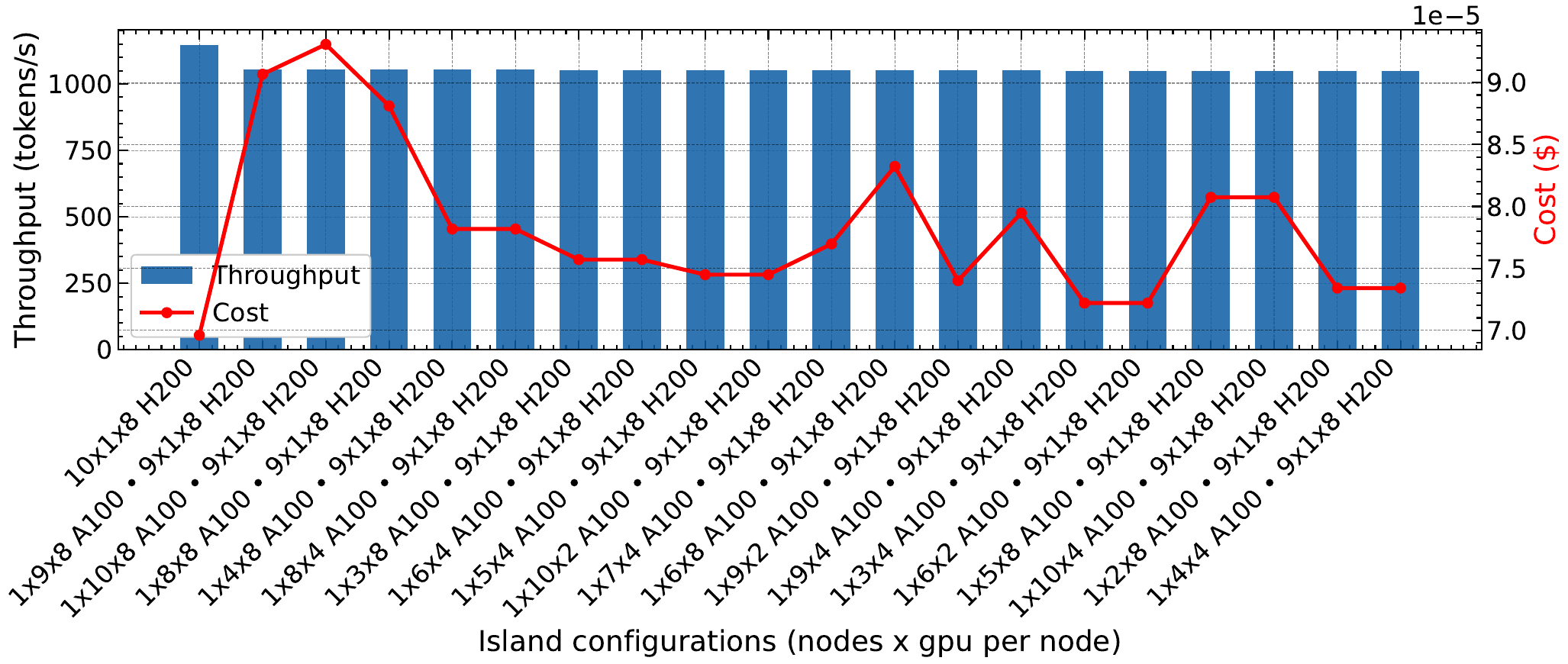}
\caption{Throughput prediction of the estimated top 20 performing configurations with optimized hyperparameters.}\label{fig:deployment_top}
\vskip -0.2in
\end{figure}
Given the assumptions outlined above, our goal is to balance training throughput and cost and demonstrate how ShardMeter facilitates this trade‑off.
Using our proposed performance model, we enumerate all feasible deployment plans of up to 10 islands across the two clusters and, for each candidate deployment, estimate both the wall‑clock time required to complete one training epoch and the corresponding monetary cost measured in GPU‑hours.
ShardMeter explicitly models the communication overhead introduced by DiLoCo‑FSDP training, ensuring that each candidate reflects a realistic deployment scenario. The reported runtime and throughput metrics, therefore, represent the final output of the performance model rather than isolated microbenchmarks.

When plotting the cost and runtime of all admissible configurations in a two‑dimensional plane, as shown in Figure~\ref{fig:deployment}, the resulting points lie along a clear Pareto frontier. Any configuration on this frontier cannot be improved in one dimension without degrading the other, indicating that no single deployment is universally optimal. Instead, the optimal choice depends on the target objective, such as maximizing throughput irrespective of cost or minimizing cost while tolerating longer wall‑clock time.
To further illustrate these trade‑offs, we select the 20 configurations on the frontier with the highest throughput and analyze their associated costs. As shown in Figure~\ref{fig:deployment_top}, throughput gains among these top configurations quickly diminish, revealing strong diminishing returns at the high‑performance end of the spectrum. In contrast, their costs vary substantially, indicating that significant cost reductions are possible even when prioritizing near‑maximal throughput.
For instance, by accepting 99\% of the maximum achievable throughput, ShardMeter identifies a deployment that reduces cost by 21.69\% relative to the throughput‑optimal configuration. More generally, for any specified performance–cost trade‑off, ShardMeter returns the corresponding island‑to‑GPU mapping, transforming an otherwise intractable design space into a set of concrete and actionable deployment decisions.
The Pareto frontier also serves as a powerful tool for selecting deployment plans that align with a user's specific budget constraints. Given a fixed cost budget, it identifies the fastest configuration. Conversely, given a time budget, it finds the least expensive configuration that meets it.

\section{Related work}
\label{sec:related_work}

A large amount of research has examined deep-learning inference and training-time prediction, but mostly focused on single-GPU jobs or conventional data-parallel/distributed training, where the compute and communication steps are straightforward and exhibit limited interdependencies. In contrast, modern large-scale sharded techniques such as FSDP and decentralized training strategies such as DiLoCo~\cite{douillard2023diloco,douillard2025streaming} introduce tightly intertwined dependencies among model partitioning, activation recomputation, and inter‑node communication, making performance prediction considerably more challenging. To the best of our knowledge, existing predictive models do not support FSDP, DiLoCo, or their derivatives. ShardMeter explicitly models and simulates these dependencies
of complex sharded or decentralized training configurations, delivering accurate runtime and monetary‑cost estimates. Its lightweight design supports exhaustive design‑space exploration, automatically uncovering deployment plans that jointly minimize execution time and expense.
We present a feature‑based overview of the related work in Table~\ref{tab:rw}.

\begin{table}[t]
\setlength{\abovecaptionskip}{-0.1in}
\caption{Comparison of existing predictors.\\(* Single node only. Trf. = Transformer architecture)}
\label{tab:rw} 
\begin{center}
\resizebox{\columnwidth}{!}{
\begin{small}
\begin{tabular}{lllclp{1cm}}
\toprule
\multirow{2}{*}{Method} & Modelling & Supported & Unseen & Parallelism & De- \\
 & Technique & DNNs & Models & Support & central \\
\midrule
Pei et al.~\cite{trainmgpu2019} & Analytical & CNN & $\checkmark$ & DP* &  -- \\
PALEO~\cite{paleo} & Analytical & CNN, GAN & $\checkmark$ & DP & -- \\
ParaDL~\cite{paradl} & Analytical & CNN & $\checkmark$ & DP, PP & -- \\ 
ConvMeter~\cite{beringer2024dissecting} & Regression & CNN & $\checkmark$ & DP & -- \\
DNNPerf~\cite{DNNPerf} & MLP & CNN, RNN & $\checkmark$ & DP* & -- \\ 
Habitat~\cite{Habitat} & MLP & CNN, Trf. & -- & DP & -- \\ 
Justus et al.~\cite{justus2018predicting} & MLP & CNN & $\checkmark$ & DP & -- \\
PreNet~\cite{pourali2024prenet} & Regr., MLP & CNN, RNN, Trf. & $\checkmark$ & DP & -- \\ 
Sailor \cite{strati2025sailor} & Analytical & Trf. & -- & DP, PP, TP & DTFM \\
\textbf{Shardmeter} & Regression & Trf. & $\checkmark$ & DP, FSDP & DiLoCo \\ 
\bottomrule
\end{tabular}
\end{small}
}
\end{center}
\vskip -0.2in
\end{table}

Pei et al.~\cite{trainmgpu2019}, PALEO~\cite{paleo}, ParaDL~\cite{paradl}, ConvMeter~\cite{beringer2024dissecting}, Justus et al.~\cite{justus2018predicting}, and DNNPerf~\cite{DNNPerf} predict training time, but they are limited to conventional DNN architectures, often only convolutional neural networks.
Habitat~\cite{Habitat} supports transformer architectures; however, its predictions are confined to a single, specific workload (e.g., a given model and batch size), and it does not cover distributed training.
PreNet~\cite{pourali2024prenet} targets transformer models by decomposing them into layers and training a separate predictor for each layer type, but its estimates are limited to data-parallel training and do not cover FSDP or decentralized strategies.
Sailor~\cite{strati2025sailor} offers a performance model for decentralized training, but it requires profiling an actual training run with their custom profiler to predict its scalability. Consequently, it can only predict the training time of known models and is unable to predict those of unknown models or training workloads.

\section{Conclusion and outlook}
\label{sec:conclusion}
In this work, we introduced ShardMeter, an analytical performance model that accurately predicts the runtime behavior of modern transformer models across diverse training regimes, including large-scale distributed and geographically decentralized settings.  Given only a model specification (e.g., architecture, size, and hyperparameters) and available hardware resources (e.g., number of nodes, GPUs per node, and network fabric), ShardMeter estimates per-GPU and per–training-island runtime and throughput without requiring prior profiling.

By explicitly modeling training execution stream and idle periods, ShardMeter identifies the root causes of performance inefficiencies and communication bottlenecks. It characterizes how batch size, model scale, GPU memory constraints, and network bandwidth jointly determine whether training is compute- or communication-bound, and pinpoints the batch-size regime in which this transition occurs. In the context of decentralized training using Streaming DiLoCo, ShardMeter further predicts when asynchronous outer-step communication becomes a bottleneck and suggests concrete mitigations. 

Beyond diagnostic insight, ShardMeter eliminates the guesswork and facilitates a cost‑aware training strategy. By enumerating feasible island‑to‑GPU mappings on heterogeneous clusters, the framework yields a spectrum of configurations that strike a balance between wall‑clock time and training cost. This trade‑off enables practitioners to select a deployment that meets a desired runtime budget while minimizing GPU‑hour cost, or vice‑versa, without exhaustively exploring an otherwise intractable design space. 
In our case study, we reduced expenses by 21\% while retaining 99\% of the maximum throughput.
Overall, ShardMeter fills a critical gap unaddressed by existing runtime predictors, which are limited to single‑GPU or homogeneous data‑parallel settings and often require prior profiling of the target model.

Looking ahead, we will broaden ShardMeter's scope to cover dynamic runtime factors, such as GPU throttling, power capping, node failures, and the impact of IO, while also making it adaptive, detecting shifts in runtime behavior and automatically adjusting its regression models on the fly.
We will extend our model to enable other parallelization methods, such as tensor parallelism, which is employed for training  models with over 100 billion parameters. 
Additionally, we will extend the cost model to include energy consumption and CO$_2$ emission estimates, enabling users to trade off runtime, monetary cost, and environmental impact when selecting a deployment configuration.

\section*{Acknowledgement}
\label{sec:conclusion}
We gratefully acknowledge support from the hessian.AI Service Center (funded by the Federal Ministry of Research, Technology and Space, BMFTR, grant no. 16IS22091), the hessian.AI Innovation Lab (funded by the Hessian Ministry for Digital Strategy and Innovation, grant no. S-DIW04/0013/003), and Germany’s Federal Ministry of Breakthrough Innovation (SPRIN-D) through the Composite Learning Challenge under the SymphonyLearn project.

\bibliographystyle{IEEEtran}
\bibliography{IEEEabrv,example_paper}

\end{document}